%% file: main_v2.tex
\documentclass[twocolumn,journal,10pt,letterpaper,final]{IEEEtran}
\usepackage[noadjust]{cite}
\usepackage{graphicx,color}
\usepackage[dvipsnames]{xcolor}
\usepackage{amsmath,amsbsy,amsfonts,amssymb}
\usepackage{mathrsfs}
\usepackage{mathtools}
\mathtoolsset{showonlyrefs}
\usepackage{subfigure}
\ifCLASSOPTIONcompsoc
\usepackage[caption=false,font=normalsize,labelfont=sf,textfont=sf]{subfig}
\else
\usepackage[caption=false,font=footnotesize]{subfig}
\fi

\usepackage{acro}
\input{acronyms.tex}

\input{notation.tex}

\begin{document}

\title{\huge An Age of Information Characterization of Frameless ALOHA}
\author{Andrea Munari, Francisco L\'azaro, Giuseppe Durisi, Gianluigi Liva
\vspace{-1.5em}
\thanks{A. Munari, F. L\'azaro and G. Liva are with the Institute of Communications and Navigation, German Aerospace Center (DLR), 82234 Wessling, Germany (email: \{andrea.munari, francisco.lazaroblasco, gianluigi.liva\}@dlr.de)\\
G. Durisi is with the Department of Electrical Engineering, Chalmers University of Technology, Gothenburg, 41296, Sweden (e-mail: durisi@chalmers.se).
}} \maketitle
\thispagestyle{empty} \setcounter{page}{0}

\maketitle

\pagestyle{empty}

\begin{abstract}
  We provide a characterization of the peak age of information (AoI) achievable in a random-access system operating according to the frameless ALOHA protocol. Differently from previous studies, our analysis accounts for the fact that the number of terminals contending the channel may vary over time, as a function of the duration of the previous contention period. The exact characterization of the AoI provided in this paper, which is based on a Markovian analysis, reveals the impact of some key protocol parameters such as the maximum length of the contention period, on the average peak AoI. Specifically, we show that setting this parameter so as to maximize the throughput may result in an AoI degradation.
\end{abstract}

\input{introduction_v2.tex}
\input{sysModel_v2.tex}
\input{framelessAloha_v2.tex}

\input{throughput_v2.tex}

\input{ageAnalysis_v2.tex}
\input{conclusions.tex}

\bibliographystyle{IEEEtran}
\bibliography{IEEEabrv,aloha}

\end{document}

%% file: acronyms.tex
\DeclareAcronym{AWGN}{short = AWGN ,long = additive white gaussian noise}
\DeclareAcronym{AoI}{short = AoI ,long = age of information}
\DeclareAcronym{PAoI}{short = PAoI ,long = peak age of information}
\DeclareAcronym{CDF}{short = CDF ,long = cumulative distribution function}
\DeclareAcronym{CRA}{short = CRA ,long = contention resolution ALOHA}
\DeclareAcronym{CRDSA}{short = CRDSA ,long = contention resolution diversity slotted ALOHA}
\DeclareAcronym{CP}{short = CP ,long = contention period}
\DeclareAcronym{CSA}{short = CSA ,long = coded slotted ALOHA}
\DeclareAcronym{C-RAN}{short = C-RAN ,long = cloud radio access network}
\DeclareAcronym{DAMA}{short = DAMA ,long = demand assigned multiple access}
\DeclareAcronym{DSA}{short = DSA ,long = diversity slotted ALOHA}
\DeclareAcronym{eMBB}{short = eMBB ,long = enhanced mobile broadband}
\DeclareAcronym{FEC}{short = FEC ,long = forward error correction}
\DeclareAcronym{GEO}{short = GEO ,long = geostationary orbit}
\DeclareAcronym{HAP}{short = HAP ,long = high-altitude platform,foreign-plural={}}
\DeclareAcronym{IC}{short = IC ,long = interference cancellation}
\DeclareAcronym{IoT}{short = IoT ,long = internet of things}
\DeclareAcronym{IRSA}{short = IRSA ,long = irregular repetition slotted ALOHA}
\DeclareAcronym{LEO}{short = LEO ,long = low Earth orbit}
\DeclareAcronym{M2M}{short = M2M ,long = machine-to-machine}
\DeclareAcronym{MAC}{short = MAC ,long = medium access}
\DeclareAcronym{MPR}{short = MPR ,long = multi-packet reception}
\DeclareAcronym{MTC}{short = MTC ,long = machine-type communications}
\DeclareAcronym{mMTC}{short = mMTC ,long = massive machine-type communications}
\DeclareAcronym{NTN}{short = NTN ,long = non-terrestrial network,foreign-plural = {}}
\DeclareAcronym{PDF}{short = PDF ,long = probability density function}
\DeclareAcronym{PER}{short = PER ,long = packet error rate}
\DeclareAcronym{PLR}{short = PLR ,long = packet loss rate}
\DeclareAcronym{PMF}{short = PMF ,long = probability mass function}
\DeclareAcronym{RA}{short = RA ,long = random access}
\DeclareAcronym{RRH}{short = RRH ,long = remote radio head,foreign-plural = {}}
\DeclareAcronym{SA}{short = SA , long = slotted ALOHA}
\DeclareAcronym{SIC}{short = SIC ,long = successive interference cancellation}
\DeclareAcronym{SIR}{short = SIR ,long = signal to interference ratio}
\DeclareAcronym{SNIR}{short = SNIR ,long = signal-to-noise and interference ratio}
\DeclareAcronym{SINR}{short = SINR ,long = signal-to-interference and noise ratio}
\DeclareAcronym{SNR}{short = SNR ,long = signal-to-noise ratio}
\DeclareAcronym{TDM}{short = TDM ,long = time division multiplexing}

%% file: notation.tex
\newcommand{\pr}{\ensuremath{\mathbb P}}
\newcommand{\prob}{\ensuremath{\mathbb P}}
\newcommand{\expOp}{\ensuremath{\mathbb E}}

\newcommand{\tru}{\ensuremath{\mathsf S}}

\newcommand{\nodes}{\ensuremath{\mathsf U}}

\newcommand{\nodesRV}{\ensuremath{U}}
\newcommand{\nodesRVl}{\ensuremath{\nodesRV^{(\ell)}}}
\newcommand{\nodesRVll}{\ensuremath{\nodesRV^{(\ell+1)}}}
\newcommand{\nodesrv}{\ensuremath{u}}

\newcommand{\maxFrame}{\ensuremath{d_{\text{max}}}}
\newcommand{\cpRV}{\ensuremath{D}}
\newcommand{\cpRVl}{\ensuremath{\cpRV^{(\ell)}}}
\newcommand{\cpRVll}{\ensuremath{\cpRV^{(\ell+1)}}}
\newcommand{\cprv}{\ensuremath{d}}

\newcommand{\decRV}{\ensuremath{M}}
\newcommand{\decRVl}{\ensuremath{\decRV^{(\ell)}}}
\newcommand{\decrv}{\ensuremath{m}}

\newcommand{\pDGivenN}{\ensuremath{P_{\cpRV|\nodesRV}}}
\newcommand{\pMGivenN}{\ensuremath{P_{\decRV|\nodesRV}}}

\newcommand{\succRV}{\ensuremath{S}}
\newcommand{\succrv}{\ensuremath{s}}
\newcommand{\succRVl}{\ensuremath{\succRV^{(\ell)}}}

\newcommand{\pActGivenL}{{\ensuremath{\mathsf \gamma_{\cprv}}}}

\newcommand{\pAct}{\ensuremath{\gamma}}
\newcommand{\pTx}{\ensuremath{q}}
\newcommand{\pUpdate}{\ensuremath{\nu}}

\newcommand{\AoI}{\ensuremath{\Delta}}

\newcommand{\pmc}[3]{\ensuremath{p_{#1}(#2,#3)}}

\newcommand{\unres}{\ensuremath{w}}
\newcommand{\collisions}{\ensuremath{c}}
\newcommand{\singletons}{\ensuremath{r}}
\newcommand{\au}{\ensuremath{i}}
\newcommand{\bu}{\ensuremath{{j}}}
\newcommand{\buall}{\ensuremath{{a}}}
\newcommand{\qu}{\ensuremath{h_\unres}}
\newcommand{\state}{\ensuremath{\mathrm{Pre}}}
\newcommand{\staten}[1]{\ensuremath{\state_{#1}}}
\newcommand{\stateafter}{\ensuremath{\mathrm{Pos}}}
\newcommand{\stateaftern}[1]{\ensuremath{\stateafter_{#1}}}
\newcommand{\probmn}[2]{\ensuremath{\beta(#1, #2)}}

\newcommand{\peakAoI}{\ensuremath{\Omega}}
\newcommand{\avPeakAoI}{\ensuremath{{\peakAoI^*}}}
\newcommand{\initAoI}{\ensuremath{\AoI_0}}
\newcommand{\initaoi}{\ensuremath{\delta_0}}
\newcommand{\interUpdate}{\ensuremath{Y}}
\newcommand{\ancMC}{\ensuremath{Z}}

\newcommand{\ancMCl}{\ensuremath{\ancMC^{(\ell)}}}
\newcommand{\ancMCll}{\ensuremath{\ancMC^{(\ell+1)}}}
\newcommand{\succRVll}{\ensuremath{\succRV^{(\ell+1)}}}

\newcommand{\decAtMaxFrame}{\ensuremath{\beta}}

\newcommand{\statDistCP}{\ensuremath{\pi_\cpRV}}
\newcommand{\statDistNodes}{\ensuremath{\pi_\nodesRV}}
\newcommand{\statDistDec}{\ensuremath{\pi_\decRV}}
\newcommand{\statDistAncMC}{\ensuremath{\pi_\ancMC}}

%% file: introduction_v2.tex
\section{Introduction} \label{se:introduction}

\Ac{IoT} systems often involve a large number of terminals that sense a physical process and report
time-stamped status updates to a common receiver. This scenario is relevant
in, e.g., environmental monitoring and asset tracking, where a primary objective is to maintain an up-to-date record of the status of an observed source.
A number of performance metrics related to the notion of information
freshness have recently been proposed to quantify the ability of a system to reach this goal~\cite{Yates21_JSAC,Uysal21_arXiv}. Among them, a prominent role is played by the
\emph{\ac{AoI}}~\cite{Kaul11_SECON}, which quantifies the amount of time elapsed since the newest
update available at the receiver was generated at the source, and has been shown to effectively capture fundamental trends in a number of relevant scenarios
\cite{Kellerer19_ACM,Sun20_TIT}.

Accurate \ac{AoI} characterizations are available for traditional grant-based link-layer medium-access policies.
Unfortunately, such policies are often highly suboptimal or non-viable in \ac{IoT} networks, because of the need to
share a common channel among a possibly massive number of battery-powered, low-complexity devices that generate traffic
in a sporadic fashion. Accordingly, random access strategies based on variations of ALOHA \cite{Abramson:ALOHA}
are the de-facto choice in a number of commercial systems \cite{LoRa,SigFox}.
Preliminary
insights on the information-freshness trade-offs that emerge in random-access systems were derived in
\cite{Yates17:AoI_SA,Yates20_ISIT}.
Specifically, these contributions illustrate that throughput and \ac{AoI} can be jointly optimized under ALOHA
policies by properly tuning the channel access probability. Further improvements in the presence of feedback were
discussed in \cite{Uysal21_JSAC,Bidokhti20_ISIT}.

In parallel to this line of research, a family of advanced grant-free protocols for \ac{IoT}, often dubbed \emph{modern
random access} \cite{Berioli16_Now,Paolini15:TIT_CSA,Stefanovic12:Frameless}, has recently been proposed. Such
protocols allow terminals to transmit multiple copies of their packets over time, and employ successive interference
cancellation at the receiver to resolve collisions.
This leads to significant throughput improvements, which makes
these solutions excellent candidates for next-generation \ac{IoT} networks. Unfortunately, little is known about the
behavior of modern random-access protocols in terms of information freshness. The first results in this direction were
presented in \cite{Munari21_TCOM}, where the focus was on irregular repetition slotted ALOHA \cite{Paolini15:TIT_CSA}.
There, non-trivial trade-offs between spectral efficiency and average \ac{AoI} were revealed.
\paragraph*{Contributions}
To further tackle this open research question, we concentrate in the present work on frameless ALOHA. This protocol,
originally proposed in \cite{Stefanovic12:Frameless}, operates according to the same principle of rateless codes, and has
emerged as a particularly promising approach. Specifically, frameless ALOHA allows terminals to transmit copies of their
packets over a contention period whose duration is dynamically tuned by the receiver. In contrast to previous works, which assume a fixed number of contending users, we focus on a more general and
realistic setup in which the number of users accessing the channel may vary over time, driven by the duration of
previous contention periods. We track the rich dynamic evolution of the system by means of a Markovian analysis,
and derive its stationary throughput. Moreover, by obtaining an exact formula for the attainable average peak \ac{AoI}, we provide the first study of the information
freshness achievable by this protocol. Our analysis highlights the critical role played
by some key protocol parameters, such as the maximum length of the contention period, and shows that operating the system to maximize throughput comes at the expense of an
\ac{AoI} degradation---a trade-off that is fundamentally different from what previously noted for traditional ALOHA strategies.

%% file: sysModel_v2.tex
\section{System Model and Preliminaries}
\label{sec:sysModel}

We focus on a system in which \nodes\ users share a wireless channel to communicate with a common receiver (sink). Time
is divided in slots of fixed duration, equal to the length of a packet, and all terminals are  slot-synchronous.
The medium is shared among all users according to a grant-free approach, and a collision channel model is assumed.
Specifically, the transmission of two or more packets over a slot leads to a destructive collision, which prevents immediate
retrieval of all colliding packets at the sink.
On the contrary, packets sent over \emph{singleton slots} are always decoded correctly.

Channel access is regulated by the frameless ALOHA protocol~\cite{Stefanovic12:Frameless}, which operates in successive
\acp{CP} of not necessarily equal length.
The receiver initiates a new \ac{CP} by broadcasting a beacon, whose duration is considered negligible throughout our analysis.
At this point,  every user with data attempts transmission of its packet over each subsequent slot with
probability~\pTx, potentially sending multiple copies of the same packet over the \ac{CP}.
Conversely, users that do not have a packet to send at the time of beacon reception refrain from accessing the channel
for the whole duration of the \ac{CP}.
The procedure continues until a new beacon sent by the sink notifies the end of the current \ac{CP} and the start of the next one.

\begin{figure*}
  \centering
  \includegraphics[width=.72\textwidth]{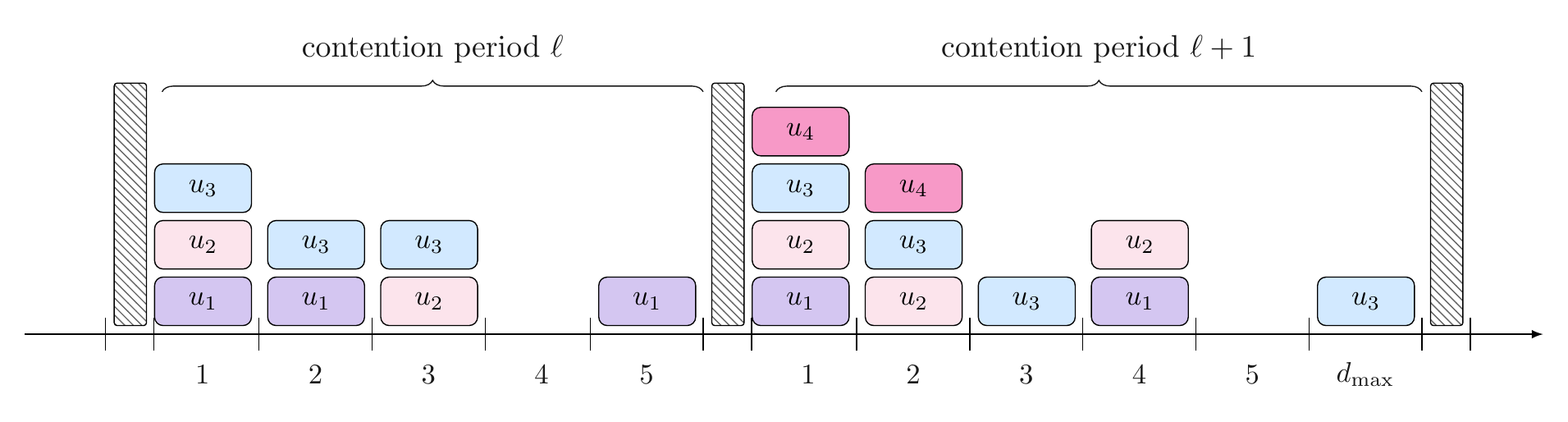}
  \caption{Example of operations for frameless ALOHA over two successive \acp{CP}.
  We assume $\nodes=4$ users in the system and a maximum contention duration of $\maxFrame=6$ slots.
Within the first \ac{CP}, only three users are active.
The receiver decodes the first packet in slot $5$, retrieving the status update of user $u_1$.
By removing its interference contribution from slot $2$, the sink can then decode the packet of user $u_3$.
Finally, after removing the interference caused by user $3$, the sink can also obtain the packet of user $2$.
Having decoded all users, the sink sends a new beacon at the end of slot $5$, initiating the next \ac{CP}.
All four users attempt transmission.
The first decoding occurs at slot $3$, leading to the retrieval of $u_3$.
Removal of such packet from slot $2$, however, does not resolve completely the existing collision, and \ac{SIC} stops.
The situation does not change after slot $4$ (collision not involving $u_3$), slot $5$ (idle), or slot $6$, which contains the transmission of a resolved user,
and the receiver terminates the \ac{CP} as the maximum number of slots has been reached,
even if some users (i.e., $u_1$, $u_2$ and $u_4$) have not been decoded.
Note that the first slot of each \ac{CP} is used by all active users to send a packet,
allowing the sink to infer when complete decoding has occurred (see Sec.~\ref{sec:operationalDetails}).}
  \label{fig:framelessTimeline}
\end{figure*}

At the receiver side, decoding of a packet over a singleton slots triggers \ac{SIC}.
Specifically, the interference contribution of all the copies of the retrieved packet is removed, possibly leading to new
singleton slots and thus to the decoding of previously collided packets.\footnote{
Note that, in order to implement this procedure, the sink needs to know of the position of all the replicas of a packet.
This can be achieved, for instance, by using a hash function of the payload as seed of a pseudo-random generator,
used by the transmitter to determine the slots of the \ac{CP} over which to transmit.
Upon decoding the payload, the sink becomes thus aware of all the slots occupied by the user, effectively allowing the removal
of the interference of that user throughout the \ac{CP}.}
The receiver proceeds with this operation mode on a slot-by-slot basis, and terminates the \ac{CP} when either all transmitting
users have been decoded or a maximum number \maxFrame~of slots has been reached.\footnote{Details on how the sink can determine whether
all users have been decoded will be presented in Sec.~\ref{sec:operationalDetails}.}
An example of the frameless-ALOHA operations is discussed in Fig.~\ref{fig:framelessTimeline}.

As to traffic generation, we assume every user to independently generate a new packet over each slot with probability~\pAct.
This packet is stored in a one-packet-sized buffer for later delivery.
A pre-emption policy with replacement in waiting is implemented, so that, at any given time instant,
a user either has one packet to send (the last generated one) or has an empty buffer.
Accordingly, a user will attempt transmission over a \ac{CP} only if it has generated at least one packet over the
previous \ac{CP}.
Assume that the previous \ac{CP} lasted for $\cprv$ slots.
Then, an arbitrary user has a packet to transmit with probability
\begin{equation}
    \pActGivenL := 1-(1-\pAct)^{\cprv}.
    \label{eq:pActGivenL}
\end{equation}
All copies of the packet sent by each user during a \ac{CP} are marked with a common time stamp, set to the start time of
the \ac{CP}.
Since all users are assumed to generate traffic independently, the number $\nodesRV$ of users that become active at the end of
a \ac{CP} of \cprv\ slots is binomial distributed with parameters $(\nodes,\pActGivenL)$.
In the remainder of the paper, we shall denote the corresponding \ac{PMF} as
\begin{align}
  P_{\nodesRV|\cpRV}(\nodesrv|\cprv) := \binom{\nodes}{\nodesrv} \pActGivenL^{\nodesrv} \, (1-\pActGivenL)^{\nodes-\nodesrv}.
  \label{eq:pUGivenD}
\end{align}
Finally, no retransmissions are considered: if a packet is not decoded during the \ac{CP} it is sent over, it is simply discarded.

\begin{figure}
    \centering
    \includegraphics[width=.82\columnwidth]{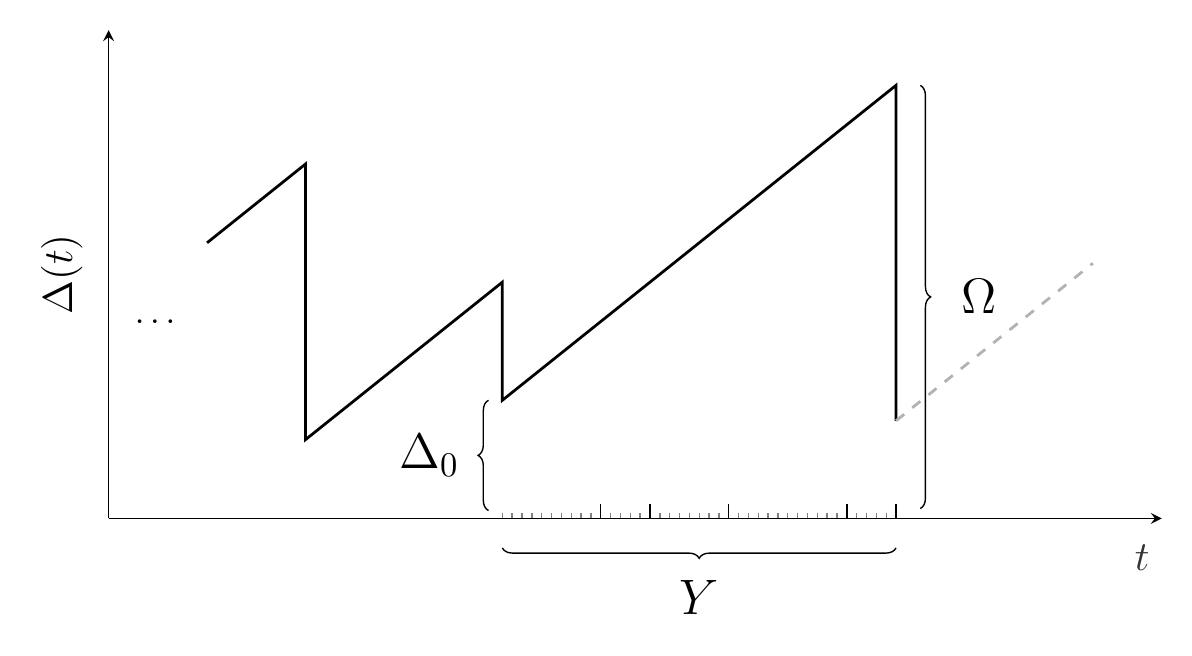}
    \caption{Evolution over time of the \ac{AoI} $\AoI(t)$ for a user.
    In the plot, \interUpdate\ denotes the duration (in slots) of an inter-update period, possibly composed by multiple \acp{CP}.
  The value at which \ac{AoI} was reset upon reception of the last update from the node is denoted by \initAoI, which in our case corresponds to the duration of the \ac{CP} in which the previous update was received.
Accordingly, \peakAoI\ indicates the value reached by $\AoI(t)$ at the end of the \ac{CP} over which the next update from the node is decoded.
An in-depth discussion of these quantities will be presented in Sec.~\ref{sec:ageAnalysis}.}
    \label{fig:aoiTimeline}
\end{figure}

We are interested in evaluating the ability of the system to maintain an up-to-date record of the state of each user at the sink.
To this aim, we consider the \ac{AoI} $\AoI(t)$ of a generic user:
\begin{equation}
\AoI(t) := t - \sigma(t).
\end{equation}
Here, $\sigma(t)$ is the time stamp of the last update received by the sink from the user of interest as of time $t$.
This metric grows linearly over time, and drops each time the receiver successfully decodes a packet from the user under
observation.
For simplicity, we will assume throughout that these refreshes take place at the end of the \ac{CP} over which the status update was received,
i.e., we do not track the exact slot in which the corresponding packet was decoded.\footnote{As will be clarified in
Sec.~\ref{sec:ageAnalysis}, this assumption does not change the fundamental trade-offs of interest, and the analysis can be easily adapted to account for this additional factor.}
This yields the saw-tooth profile exemplified in Fig.~\ref{fig:aoiTimeline}.

We will focus on the \emph{average peak \ac{AoI}} \avPeakAoI, defined as the mean value of $\AoI(t)$ when sampled right before an update
from the user of interest is decoded.
This metric, introduced in \cite{Ephremides14:peakAge}, characterizes the maximum value reached on average by the
\ac{AoI} of a user.

\input{sysModel_Remarks_v2.tex}
\subsection{Notation}
In the remainder of the paper, we denote a discrete r.v. and its realization as $X$ and $x$, respectively, whereas the corresponding \ac{PMF} is indicated by $P_X(x)$. A conditional \ac{PMF} is denoted as $P_{X|Y}(x|y)$. We further write the state of a discrete-time Markov chain at time $\ell$ as $X^{(\ell)}$, and express its one-step transition probability from state $i$ to state $j$ as
\begin{equation*}
    \pmc{X}{i}{j} := \mathbb P \{X^{(\ell+1)} = j \, | \, X^{(\ell)} = i\}\,.
\end{equation*}
In the case of bi-dimensional Markov chains, we maintain the same notation, but denote the state by means of a
two-element vector, e.g., $j = (j_1,j_2)$.

%% file: sysModel_Remarks_v2.tex
\subsection{Operational Details}
\label{sec:operationalDetails}

We next describe some operational details of the protocol that will be relevant for the subsequent analyses.
At the end of each slot, the sink attempts to decode as many users as possible, canceling also their interference.
When no more users can be decoded, i.e., when the contention contains no more singleton slots, the receiver decides
whether to terminate the \ac{CP} or not.
Specifically, the \ac{CP} is concluded only if all active users have been decoded, or alternatively
if a maximum number of slots has elapsed since the beginning of the contention.
Note, however, that, without further assumptions, it is in general not possible for the sink to determine whether all active users have been decoded,
since the sink cannot discriminate between inactive users, who do not have a packet to transmit,
and active users who do have a packet to transmit, but have not (yet) transmitted their packet since the beginning of the \ac{CP}.

To allow the sink to determine whether all active users have been decoded, we set the slot access probability to $1$ in the first slot of every contention period.
This implies that all active users will transmit their packet in the first slot.
Furthermore, we make the reasonable assumption that the receiver can distinguish among {empty slots}, {singleton slots} containing exactly one packet, and \emph{collided slots} containing two or more packets.
Under this assumption, the sink can use the first slot of every \ac{CP} to determine whether all active users have been decoded or not.
In particular, after canceling the interference from a decoded user, the sink can check whenever the first slot becomes
empty to infer whether there are no more undecoded active users and the \ac{CP} can be terminated.
This strategy allows the receiver also to detect empty \acp{CP}.
Indeed, these \acp{CP} are characterized by an empty initial slot.
Note that the minimum \ac{CP} duration in our setting is one slot, reached when either no users or a single user have data to transmit.

We emphasize that more realistic and sophisticated methods may be devised to estimate the number of active users in the \ac{CP}, as discussed for instance in~\cite{Stefanovic13:RatelessAloha}. For the purpose of the analysis provided in this paper, the proposed technique suffices, in the sense that it provides a simple model for the cost (i.e., overhead) required for the estimation of the number of active users.

%% file: framelessAloha_v2.tex
\section{Frameless ALOHA Analysis}
\label{sec:framelessAnalysis}

Following \cite{Lazaro20_TCOM}, we model the iterative \ac{SIC} process at the sink using a finite-state machine.
A state is identified by the triplet $( \unres, \collisions, \singletons)$, where $\unres$ denotes the number of unresolved users, 
$\collisions$ denotes the number of collided slots (ignoring the initial slot),  and $\singletons$ is the number of singleton slots.
We denote by $\staten{\cprv}$ the pre-decoding state, i.e., the state right after the sink observes the $\cprv$-th slot within a \ac{CP} and before it tries to decode any new packets, whereas $\stateaftern{\cprv}$ denotes the post-decoding state, i.e., the state after \ac{SIC} decoding.
To describe the decoding process, we next provide a characterization of the conditional probability of $\stateaftern{\cprv}$ given $\staten{\cprv}$ and
of the conditional probability of $\staten{\cprv}$  given  $\stateaftern{\cprv-1}$.

\subsection{State Initialization}
Assume that $u$ users are active. 
The state is initialized as 
\[
\staten{1}= \begin{cases}
	( 0,  0, 0)  & \text{if } \nodesrv=0	\\
	( 1,  0, 1)  & \text{if } \nodesrv=1   \\
	( \nodesrv,  0, 0)  & \text{if } \nodesrv\geq 1.
\end{cases}
\]

\subsection{Conditional Probability of $\stateaftern{\cprv}$ Given $\staten{\cprv}$}
We next derive the conditional probability of the post-decoding state $\stateaftern{\cprv}$ given the pre-decoding state 
$\staten{\cprv} = ( \unres,  \collisions, \singletons)$.
Two cases need to be distinguished: $\singletons=0$ and $\singletons=1$.
Indeed, in the pre-decoding state, we always have $r\in\{0,1\}$, since the reception of a new slot yields at most one new singleton slot. 
If $\singletons=0$, the state remains unchanged since no users can be resolved. 
Hence, we have 
\begin{align}
	&\prob\{ \stateaftern{\cprv} = ( \unres',  \collisions', \singletons')| \staten{\cprv} = ( \unres, \collisions, 0)\}  \\
	&   \qquad \qquad \qquad
	= \begin{cases}
		1 & \text{if } \unres'=\unres, \,   \collisions' = \collisions, \, \singletons' = 0 \\
		0 & \text{otherwise}.
	\end{cases}
\end{align}

Let us now focus on the case $\singletons=1$.
It is convenient to describe \ac{SIC} decoding as an iterative process in which one user is resolved at a each iteration, potentially resulting in new singleton slots. 
The iterative process is terminated when no singleton slots are available.
This implies that the post-decoding state must have $\singletons=0$.
To characterize the state evolution at each \ac{SIC} iteration, we use~\cite[Theorem 1]{Lazaro20_TCOM}.
This theorem, when specialized to the scenario considered here, implies that, if
the state is $(\unres,  \collisions, \singletons)$ with $\unres\geq1$ and $\singletons\geq1$, after resolving exactly one user, 
the state becomes $(\unres-1,  \collisions-\bu, \singletons-\au+\bu+\buall)$ with probability
\begin{equation}
I_\unres(\buall) \binom{\collisions}{\bu} \qu^\bu (1-\qu)^{\collisions-\bu} \binom{\singletons-1}{\au-1} \left( \frac{1}{\unres}\right)^{\au-1} \left(1- \frac{1}{\unres}\right)^{\singletons-\au}
\end{equation}
for $0\leq \buall \leq 1$, $0\leq \bu \leq \collisions$, $1 \leq \au \leq \singletons-1$ and $\au -\bu -\buall \leq \singletons$, 
where 
\begin{equation}
I_\unres(\buall)=\begin{cases}
	1 & \text{if } \unres\neq2, \, \buall=0 \\
	1 & \text{if } \unres=2, \, \buall=1 \\
	0 & \text{otherwise}
\end{cases}
\end{equation}
and with 
\begin{equation}
\qu =\frac{\sum\limits_{k=2}^{\nodesrv-\unres+2} \Lambda_k k (k-1) \frac{1}{\nodesrv} \frac{\unres-1}{\nodesrv-1} \frac{\binom{\nodesrv-\unres}{k-2}}{\binom{\nodesrv-2}{k-2}}} {1 - \sum\limits_{k=1}^{\nodesrv-\unres+1} \Lambda_k \unres \frac{\binom{\nodesrv-\unres}{k-1}}{\binom{\nodesrv}{k}} - \sum\limits_{k=0}^{\nodesrv-\unres} \Lambda_k  \frac{\binom{\nodesrv-\unres}{k}}{\binom{\nodesrv}{k}} }
\end{equation} 
where $\Lambda_k=\binom{\nodesrv}{k} \pTx^k (1-\pTx)^{\nodesrv-k}$.

To derive the desired conditional probability $\prob\{ \stateaftern{\cprv} = ( \unres', \collisions', 0)| \staten{\cprv} = ( \unres,  \collisions, 1)\}$ 
for all values of $\unres'$, and $\collisions'$, we apply the result just stated iteratively, stopping when we reach a state with no singleton slots.

\subsection{Contention Termination}
The \ac{CP} is terminated after $\cprv<\maxFrame$ slots only if all $\nodesrv$ active users are resolved, i.e., only if the post-decoding state is $\stateaftern{\cprv} =  (0, 0, 0)$.
However, when $\cprv=\maxFrame$, the \ac{CP} is terminated, no matter what the value of $\stateaftern{\maxFrame}$ is.

\subsection{Conditional Probability of $\staten{\cprv}$  Given  $\stateaftern{\cprv-1}$}
We now analyze how the state changes when one slot is added to the \ac{CP}.
To do so, we derive the conditional probability of the pre-decoding state $\staten{\cprv}$ given the post-decoding state
$\stateaftern{\cprv-1}=( \unres, \collisions, 0)$, for $\cprv \geq 2$. 
Three different cases must be considered. 
In the first case, the extra slot contains no packet from any of the $\unres$ unresolved users. 
This event, which occurs with probability $(1-\pTx)^{\unres}$, yields a pre-decoding state 
$\staten{\cprv} = ( \unres, \collisions, 0)$. 
Hence, we have
\[
\prob \{ \staten{\cprv} = ( \unres, \collisions, 0)| \stateaftern{\cprv-1} = ( \unres,  \collisions, 0)\} = (1-\pTx)^{\unres}.
\]
In the second case, the extra slot contains the packet of exactly one of the $\unres$ unresolved users.
It can then be verified that
\[
\prob \{ \staten{\cprv} = ( \unres,  \collisions, 1)| \stateaftern{\cprv-1} = ( \unres,  \collisions, 0)\} =  \unres \pTx (1-\pTx)^{\unres-1}.
\]
Finally, in the third case, the extra slot contains the transmission of two or more unresolved users, which yields  
\begin{align}
\prob \{ \staten{\cprv} = ( \unres,  \collisions+1, 0)| \stateaftern{\cprv-1} = ( \unres,  \collisions, 0)\} &=  
1- (1-\pTx)^{\unres} \\
&\mkern-30mu\phantom{=}- \unres \pTx (1-\pTx)^{\unres-1}.
\end{align}
\subsection{Derivation of Some Useful Quantities}
We will next use the state-transition probabilities just introduced to derive three main quantities that will turn out important
for the characterization of the average peak \ac{AoI}.

The first quantity is the probability that the contention period is terminated after exactly $\cprv$ slots, given that
the number of active users is $\nodesrv$. We denote this quantity by $\pDGivenN( \cprv | \nodesrv)$. 
To characterize it, we need to consider three different cases.
The first one is $\cprv=1$. In this case, we have
\[
\pDGivenN( 1 | \nodesrv) = \begin{cases}
	1 &\text{if } \nodesrv\leq1 \\
	0 & \text{otherwise}.
\end{cases}  
\]
The second case covers ${1 <\cprv<\maxFrame}$.
Recall that, in this case, the \ac{CP} is terminated only if all $\nodesrv$ active users are resolved.  
Hence, 
\begin{align}
\pDGivenN( \cprv | \nodesrv) = \prob \{ \stateaftern{\cprv}= ( 0, 0, 0)  \}.
\label{eq:pDGivenU_a}
\end{align}
The probability of the  remaining case, $\cprv=\maxFrame$, can be easily obtained as
\begin{align}
	\pDGivenN( \maxFrame | \nodesrv) 
	&= 1 - \sum_{\cprv=1 }^{\maxFrame-1} \pDGivenN( \cprv | \nodesrv).
\label{eq:pDGivenU_b}
\end{align}

The second quantity we are interested in is the conditional probability that exactly $m$ users were decoded given that $\nodesrv$
users were active.
We denote this quantity by $\pMGivenN(\decrv|\nodesrv)$. 
To characterize it, we must distinguish two cases: $m<\nodesrv$ and $m=\nodesrv$. 
When $m<\nodesrv$, since not all users were resolved, the \ac{CP} was terminated after $\maxFrame$ slots. 
Hence, we have
\[
\pMGivenN(\decrv|\nodesrv)=  \sum_{\collisions} \prob \{ \stateaftern{\maxFrame}= ( \nodesrv-m, \collisions, 0) \}.
\]
Assume now $m=\nodesrv$.
To obtain $\pMGivenN(\nodesrv|\nodesrv)$, we need to add the probabilities of all post-decoding states in which all users are decoded:
\begin{IEEEeqnarray}{Clr}
\pMGivenN(\nodesrv|\nodesrv)& =& \sum_{\cprv=1}^{\maxFrame} \prob \{ \stateaftern{\cprv}= ( 0, 0, 0) \}\notag\\
&=& 1 - \sum_{\decrv=0 }^{\nodesrv-1} \pMGivenN(\decrv|\nodesrv). \notag
\end{IEEEeqnarray}

The third quantity of interest, which we denote by $\probmn{m}{\nodesrv}$, is the conditional probability that $m$ users are
resolved, given that $\nodesrv$ users accessed the \ac{CP} and that the \ac{CP} ran until its maximum duration $\maxFrame$. 
We can obtain $\probmn{m}{\nodesrv}$ by summing  the probabilities of all post-decoding states $\stateaftern{\maxFrame}$
in which exactly $\nodesrv-m$ active users are unresolved, and then normalizing by the sum of the probabilities of all states $\stateaftern{\maxFrame}$:
\begin{align}
\probmn{m}{\nodesrv} = \frac{\sum_{\collisions} \prob \{ \stateaftern{\maxFrame}= ( \nodesrv-m,  \collisions, 0) \} }{\sum_{\unres}  \sum_{\collisions} \prob \{ \stateaftern{\maxFrame}= ( \unres, \collisions, 0) \}}.
\label{eq:probDecAtMaxFrame}
\end{align}


%% file: throughput_v2.tex
\section{Throughput Performance}
\label{sec:throughput}
We provide in this section an analysis of the stationary throughput achievable with the frameless ALOHA protocol.
This analysis will turn out useful for the characterization of the \ac{AoI}.
 Previous works, e.g., \cite{Stefanovic12:Frameless,Lazaro20_TCOM,Stefanovic13:RatelessAloha,Stefanovic13_ICC}, have studied the protocol
 behavior either over a single \ac{CP}, or under the assumption that the number of contending terminals is fixed. 
 For this scenario, the number of packets that can be decoded under an optimized access probability has been
 characterized.
 The setting under consideration in this paper, however, is characterized by a richer dynamic, since the number of 
 users with packets to transmit, and thus the level of contention, may vary over time. 
 To appreciate this aspect, observe how, for instance, a long \ac{CP} increases the probability for more users to
 generate at least one packet over its duration. 
 This leads to a harsher contention over the successive period, which, in turn, is likely to last longer. 
 Similarly, contentions resolved in few slots will instead drive the system on average towards shorter 
 and less loaded \acp{CP}.

To capture the impact on throughput of this non-trivial evolution, we focus on the homogeneous Markov processes \cpRVl\
and \nodesRVl, tracking the duration of the $\ell$-th \ac{CP} and the number of users contending over it, respectively.
Let us first consider the former, which takes values in the set $\{1,\dots,\maxFrame\}$. Recalling that the duration of
the $(\ell+1)$-th \ac{CP} is driven by the number of users contending over it, we compute the transition probabilities for the
chain as
\begin{align}
  p_{\cpRV}(i,j) &= \sum_{\nodesrv=0}^\nodes \pr\{\cpRVll=j \,|\, \nodesRVll = \nodesrv\}\\      &\qquad \times\pr\{\nodesRVll = \nodesrv \,|\, \cpRVl = i\}\\
  &\stackrel{(a)}{=}\sum_{\nodesrv=0}^\nodes P_{\cpRV | \nodesRV} (j|\nodesrv) \, P_{\nodesRV|\cpRV}(\nodesrv|i)
  \label{eq:transMC_D}
\end{align}
where (a) follows from~\eqref{eq:pUGivenD}, \eqref{eq:pDGivenU_a}, and~\eqref{eq:pDGivenU_b}.
Similarly, the transitions for the finite-state chain \nodesRVl\ are
\begin{align}
  p_{\nodesRV}(i,j) &= \sum_{\cprv=1}^{\maxFrame} \pr\{\nodesRVll=j \,|\, \cpRVl = \cprv\}\\      &\qquad \times\pr\{\cpRVl = \cprv \,|\, \nodesRVl = i\}\\
  &=\sum_{\cprv=1}^{\maxFrame} P_{\nodesRV | \cpRV} (j|\cprv) \, P_{\cpRV|\nodesRV}(\cprv|i).
  \label{eq:transMC_U}
\end{align}
In both cases, it is easy to verify that the finite-state chains are irreducible and aperiodic, and thus ergodic. In the remainder of the discussion, we shall indicate their stationary distributions, derived by solving the corresponding balance equations, as $\statDistCP(\cprv)$ and $\statDistNodes(\nodesrv)$, respectively.

Let us now denote by \decRVl\ the number of successfully decoded users over the $\ell$-th \ac{CP}. 
The system throughput \tru, defined as the \emph{average} number of decoded packets per slot, is
\begin{align}
  \tru := \lim_{t\rightarrow\infty} \frac{\frac{1}{t} \sum_{\ell=1}^t \decRVl}{\frac{1}{t} \sum_{\ell=1}^t \cpRVl}.
  \label{eq:truDef}
\end{align}
Observing now that 
\begin{align}
\pr\{\decRVl = \decrv\} = \sum_{\nodesrv=0}^\nodes P_{\decRV|\nodesRV}(\decrv|\nodesrv) \, \pr\{\nodesRVl=\nodesrv\}
\end{align}
we conclude that the statistics of \decRVl\ can be directly derived from that of the number of contending users over the
corresponding \ac{CP}. Hence, this process has also a stationary distribution.
Accordingly, both numerator and denominator in~\eqref{eq:truDef} admit finite limits for $t\rightarrow\infty$ by virtue
of the ergodicity of the involved chains. This allows us to to compute \tru\ as the ratio of the expected values of the processes in stationary conditions: 
\begin{align}
  \vspace{.3em}
  \tru = \frac{\sum_{\decrv = 0}^\nodes \sum_{\nodesrv=0}^\nodes \decrv \, P_{\decRV|\nodesRV}(\decrv|\nodesrv) \, \statDistNodes(\nodesrv)}{\sum_{\cprv=1}^{\maxFrame} \cprv \,\statDistCP(\cprv)}.
  \vspace{.3em}
  \label{eq:tru}
\end{align}

\begin{figure}
  \centering
  \includegraphics[width=0.85\columnwidth]{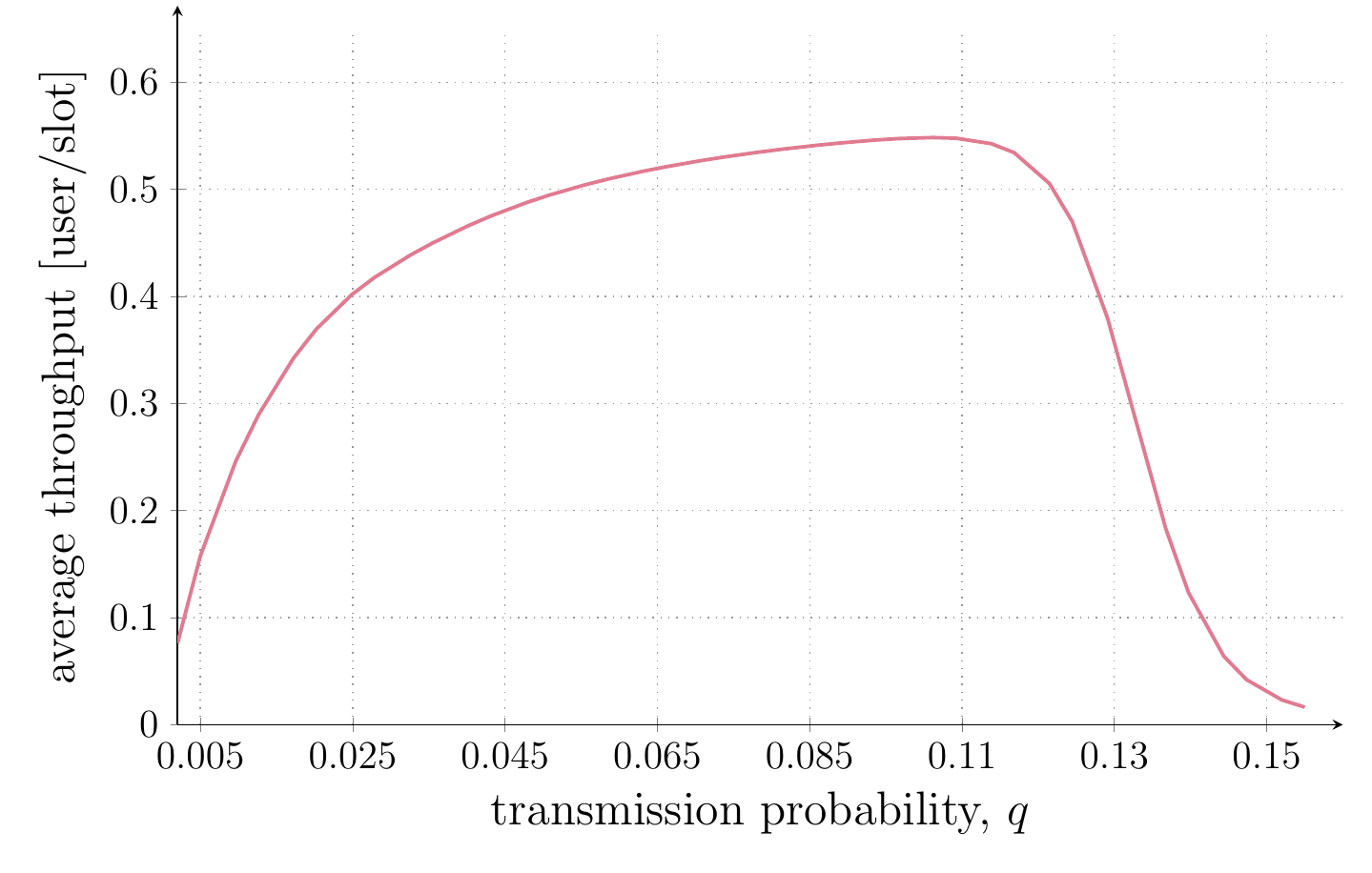}
  \caption{System throughput \tru\ vs transmission probability \pTx. A population of $\nodes=100$ users and a maximum
  \ac{CP} duration of $\maxFrame=100$ slots are considered. The packet generation probability is set so that $\pAct \nodes = 0.6$.}
  \label{fig:truVsPtx}
\end{figure}

Leaning on this result, we provide a first characterization of the behavior of the system in Fig.~\ref{fig:truVsPtx}. 
The plot shows how the stationary throughput changes as a function of the transmission probability \pTx, considering a
population of $\nodes=100$ users, and a maximum duration for a \ac{CP} of $\maxFrame=100$ slots. The reported results
were obtained by setting the activation probability \pAct\ such that the average number of users generating a new packet
over each slot, $\pAct \nodes$, equals $0.6$. 
The exhibited trend confirms the existence of an optimal medium access probability for throughput performance. 
Indeed, too low values of \pTx\ tend to result in successful yet unnecessarily long \acp{CP}, where many slots may
remain unused. Conversely, when users become too aggressive, collisions become predominant, leading to the sharp
decrease in throughput which is typically observed in grant-free schemes that resort to \ac{SIC}.  

\begin{figure*}[t]
  \begin{center}
  \includegraphics[width=.165\textwidth]{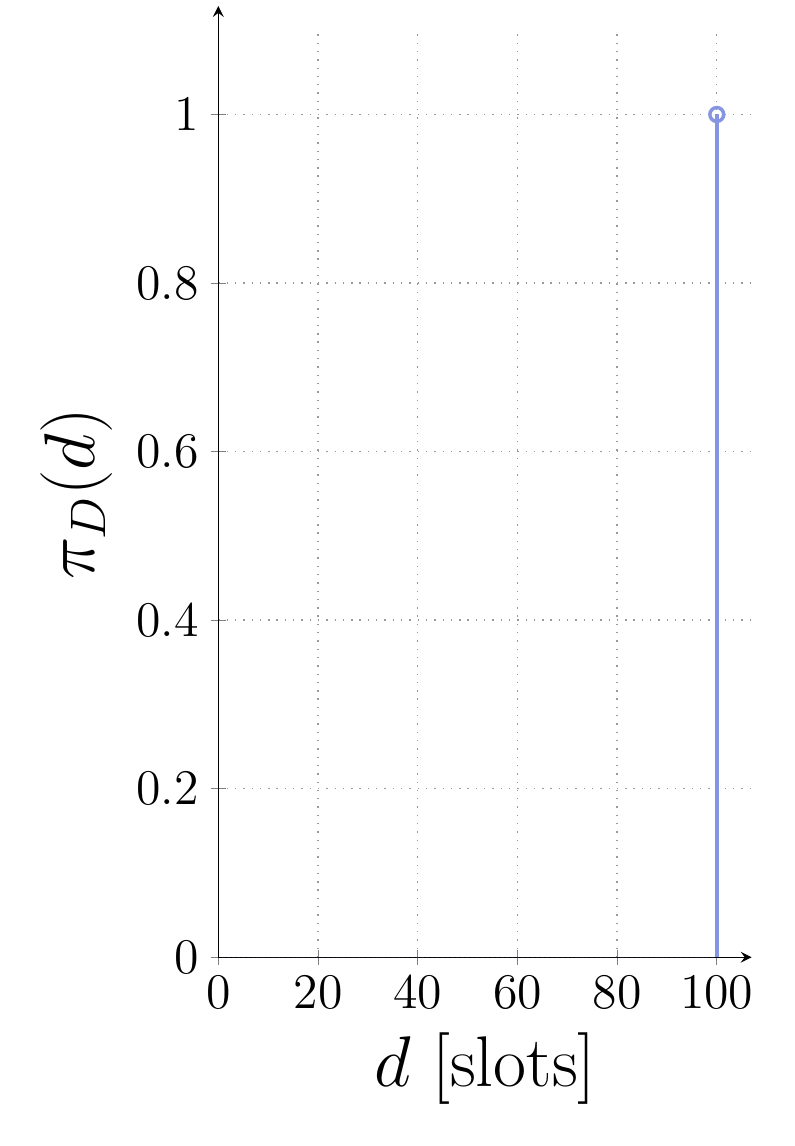}
  \hspace{-1em}
  \includegraphics[width=.165\textwidth]{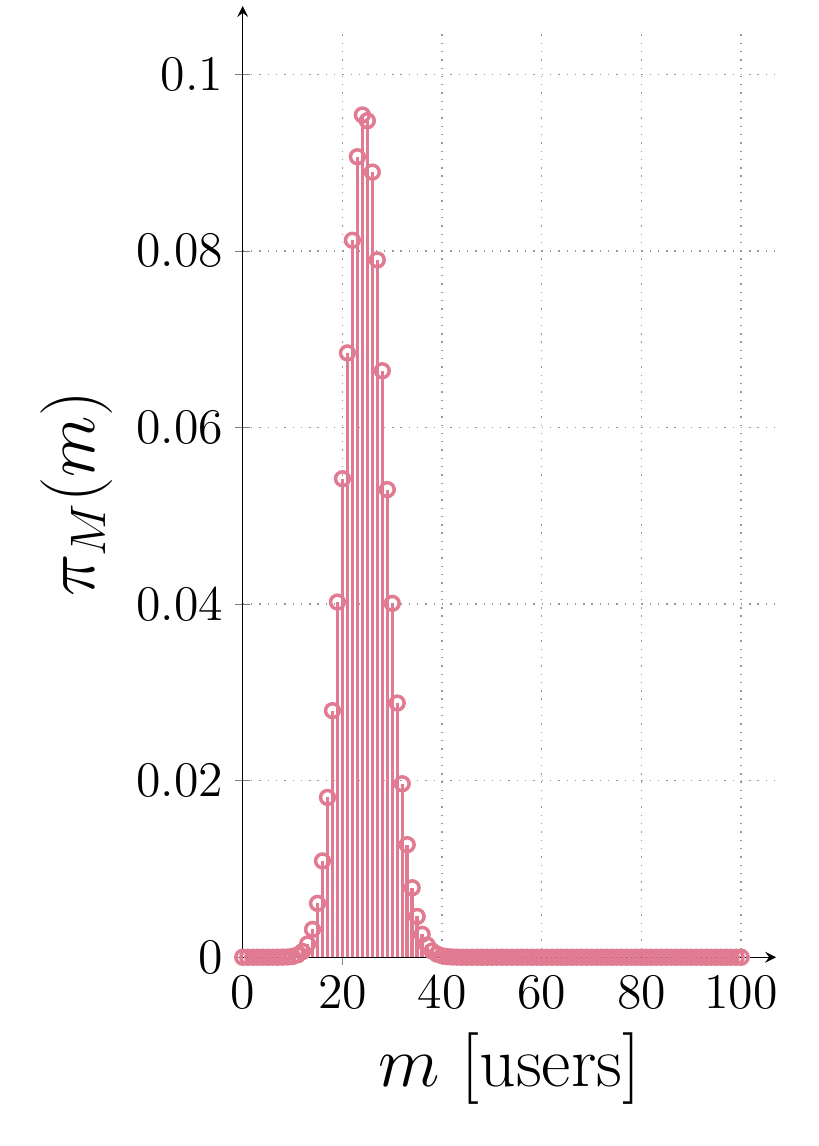}
  \includegraphics[width=.165\textwidth]{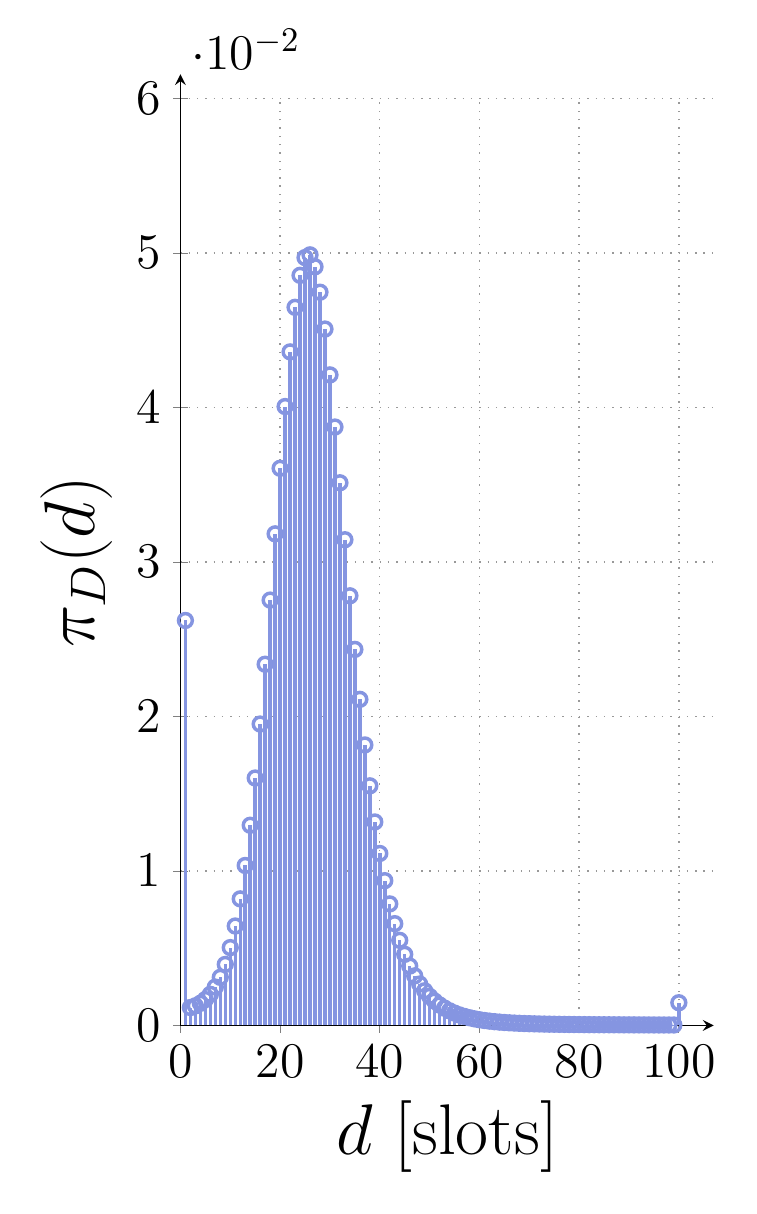}
  \hspace{-1em}
  \includegraphics[width=.165\textwidth]{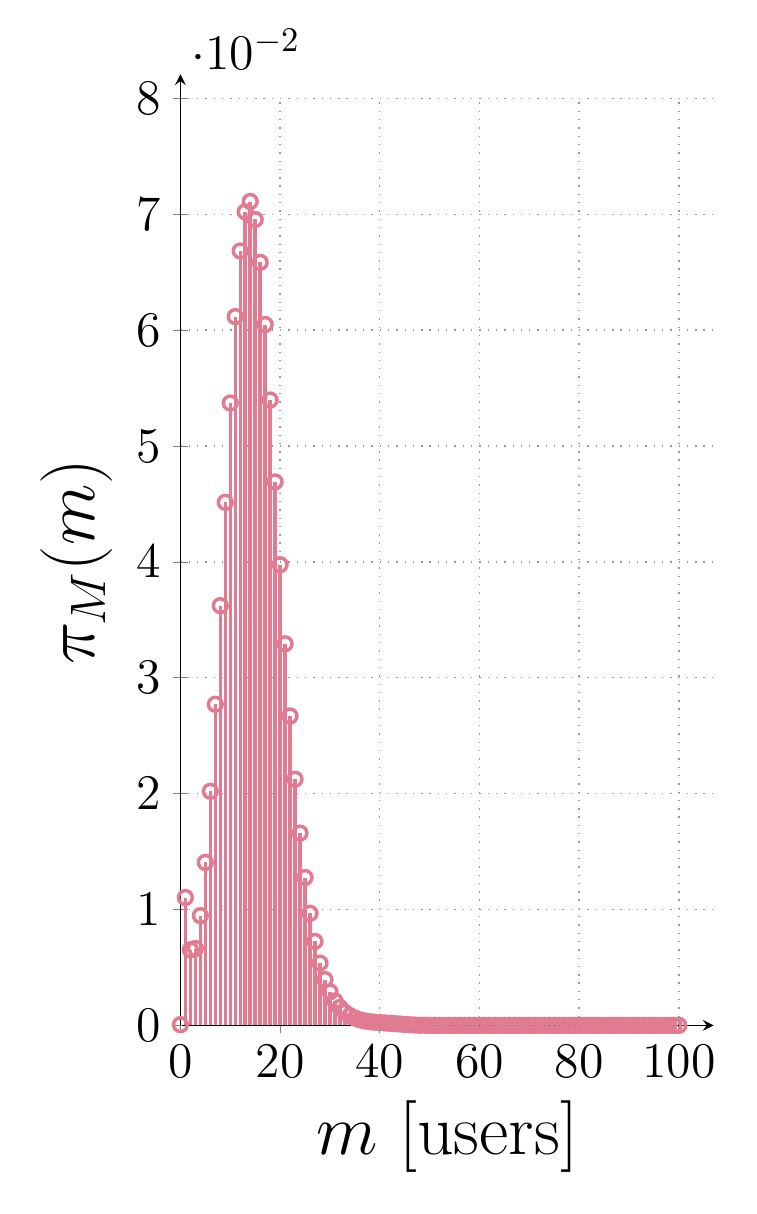}
  \includegraphics[width=.165\textwidth]{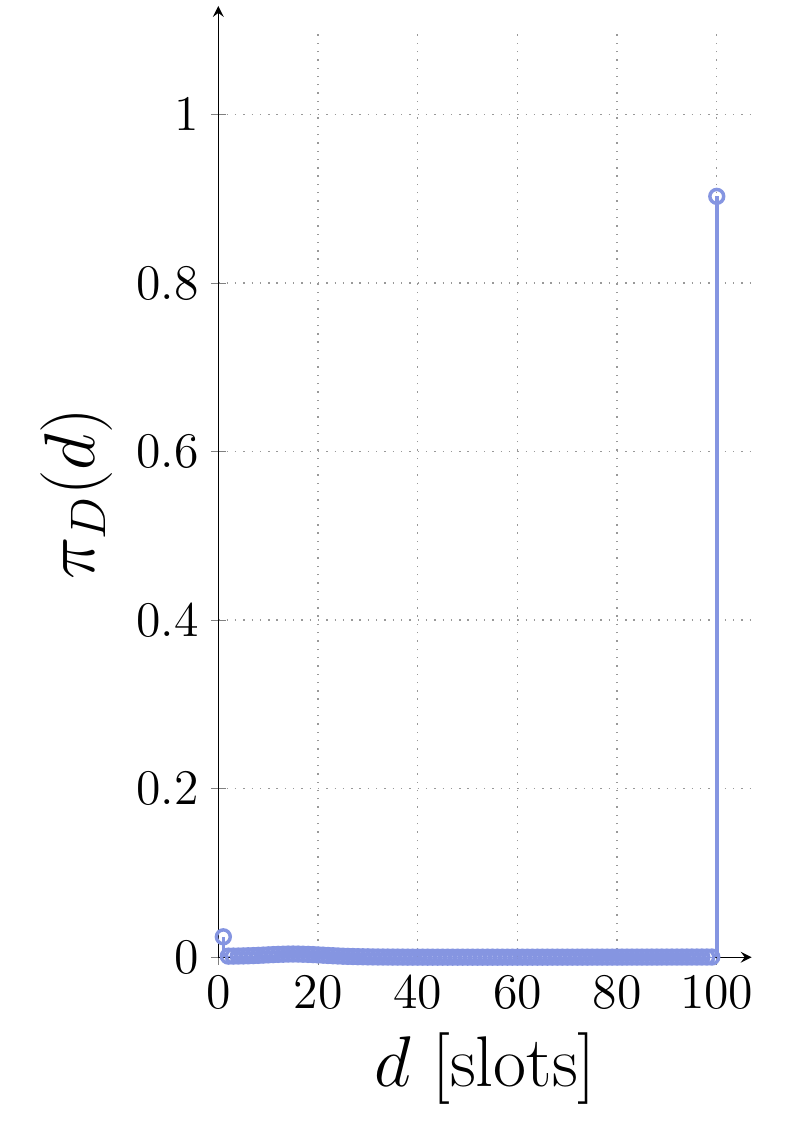}
  \hspace{-1em}
  \includegraphics[width=.165\textwidth]{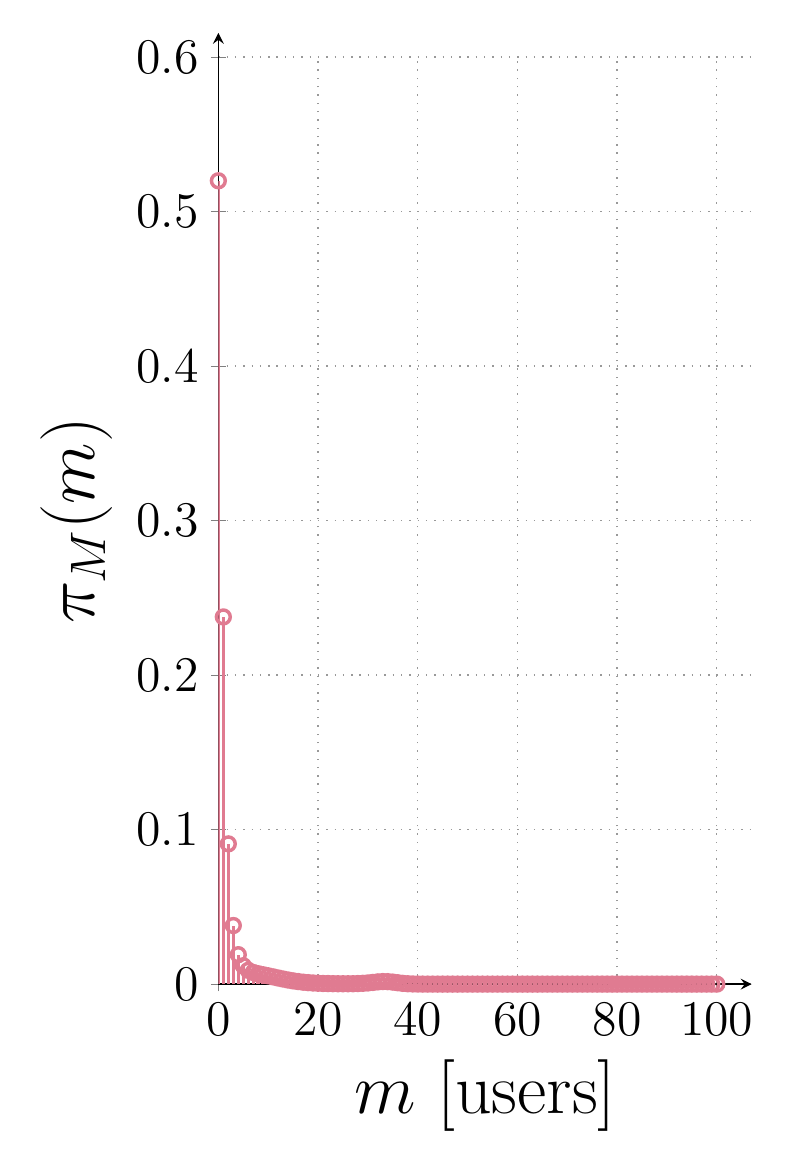}
  \end{center}
  \vspace{-.6em}
  \footnotesize
  \hspace{9em} $(\pTx=0.01)$ \hspace{16em} $(\pTx=0.1)$ \hspace{17em} $(\pTx=0.15)$
  \caption{Stationary distribution of the \ac{CP} duration, $\statDistCP(\cprv)$, and of the number of decoded users per \ac{CP}, $\pi_{\decRV}(\decrv)$, for three different values of transmission probability $\pTx$. In all cases, results were generated considering $\nodes=100$, $\maxFrame=100$, and $\pAct \nodes = 0.6$. The \emph{average} number of active users per \ac{CP} in stationary conditions in the different cases is $45.22$ for $q=0.01$, $14.49$ for $q=0.1$, and $41.68$ for $q=0.15$. }
  \label{fig:statDistrib}
\end{figure*}

Such a behavior is confirmed in Fig.~\ref{fig:statDistrib}, which shows the stationary distribution of the \ac{CP}
duration, $\statDistCP(\cprv)$, and of the number of decoded packets per \ac{CP}, denoted as $\statDistDec(\decrv)$, for
low ($\pTx=0.01$), intermediate ($\pTx=0.1$) and high ($\pTx=0.15$) values of the transmission probability. Consider
first the case $\pTx=0.01$. In this situation, an active node will send no copy of its packet even over a \ac{CP} of
maximum duration with probability $(1-\pTx)^{\maxFrame} \simeq 0.37$. As a result, the sink tends to operate with long
\acp{CP}, awaiting for packets of users that have not yet transmitted. In terms of throughput, while an average of
around $45$ users will participate in the contention for a \ac{CP} of $\maxFrame=100$ slots, only a significantly lower
fraction of them is decoded, as reported by the stationary distribution $\statDistDec(\decrv)$. Notably, the system
operates at \acp{CP} of maximum duration also for high values of transmission probability ($\pTx=0.15$). In this case,
though, hardly any packet is decoded, due to the excessive level of congestion, which leads to unresolvable collisions. Both
settings result in poor throughput performance, as illustrated in Fig. \ref{fig:truVsPtx}. An efficient utilization of
the channel is instead achieved for $\pTx = 0.1$, where a proper balance between frame duration and number of active users
is reached. A first relevant trade-off driven by the dynamics of the system thus emerges, calling for a proper balance
of the operating parameters. 

%% file: ageAnalysis_v2.tex
\section{Average Peak Age of Information}
\label{sec:ageAnalysis}

We now analyze the performance of frameless ALOHA in terms of information freshness. We start with
some preliminary results that will facilitate the characterization of the average peak \ac{AoI} \avPeakAoI.

\subsection{Preliminaries}
Fix a generic user for which the \ac{AoI} is tracked, and denote by $\pUpdate(\nodesrv,\cprv)$ the conditional probability
that the user delivers a status update over the current \ac{CP}, given that $\nodesrv$ users contend,
and that the \ac{CP} is terminated after $\cprv$ slots. Recall that whenever the
\ac{CP} is terminated prior to its maximum duration, all contending users are successfully decoded.
The user of interest belongs to this pool with probability $\nodesrv/\nodes$. Conversely, if the \ac{CP} runs for
\maxFrame\ slots, the conditional probability for the user to deliver a packet given that $\decrv$ users are
successfully decoded is $(\decrv/\nodes)\decAtMaxFrame(\decrv,\nodesrv)$.
Combining these two results we then have:
\begin{align}
    \pUpdate(\nodesrv,\cprv) =
    \begin{dcases}
    \,\,\frac{\nodesrv}{\nodes} & \quad \cprv < \maxFrame \\
    \,\,\sum_{\decrv=0}^\nodesrv \frac{\decrv}{\nodes} \,\, \beta(\decrv,\nodesrv) & \quad \cprv = \maxFrame.
    \end{dcases}
    \label{eq:pup}
\end{align}

Next, we introduce a simple ancillary Markov chain, whose state is defined as \mbox{$\ancMCl = (\cpRVl, \succRVl)$}. The
first component, which we have already discussed, characterizes the duration of the $\ell$-th \ac{CP}, whereas \succRVl\ is a binary
r.v. taking value $1$ if an update from the user of interest has been successfully received over the $\ell$-th \ac{CP},
and $0$ otherwise. Consider now the probability for the chain to transition from state $(j,\succrv)$ to state
$(\cprv,1)$. By definition, this event occurs when the current \ac{CP} has duration $\cprv$ slots, and the user
delivers an update. Observing that the user's success does not depend on its outcome over the previous \ac{CP}, we can
simplify the transition probability to
\begin{align}
      p_{\ancMC}((j,\succrv),(\cprv,1)) &:=\pr\{ \ancMCll = (\cprv,1) \,|\, \ancMCl = (j,\succrv) \}\\
    &= \pr\{ \succRVll = 1, \cpRVll = \cprv \,|\, \cpRVl = j \}.
\end{align}
Conditioning now on the number of users contending over the \ac{CP}, we further have
\begin{align}
  \pr\{ &\succRVll = 1, \cpRVll = \cprv \,|\, \cpRVl = j \} \\
  &= \sum_{\nodesrv=0}^\nodes \pr\{\succRVll=1 \,|\, \nodesRVll=\nodesrv, \cpRVll=d\}\\
  &\quad   \times\pr\{ \nodesRVll=\nodesrv, \cpRVll=\cprv \,|\, \cpRVl = j\}\\
  &= \sum_{\nodesrv=0}^\nodes \pr\{\succRVll=1 \,|\, \nodesRVll=\nodesrv, \cpRVll=d\} \\
  &\quad \times \pr\{\cpRVll\!\!=\cprv \,|\, \nodesRVll\!\!=\nodesrv \} \, \pr\{\nodesRVll\!\!=\nodesrv \,|\,\cpRVl \!\!= j\}.
\end{align}
Finally, using~\eqref{eq:pup},~\eqref{eq:pUGivenD}, \eqref{eq:pDGivenU_a}, and~\eqref{eq:pDGivenU_b}, we can write
$p_{\ancMC}((j,\succrv),(\cprv,1))$ compactly as
\begin{equation}
     p_{\ancMC}((j,\succrv),(\cprv,1)) =\sum_{\nodesrv=0}^\nodes \pUpdate(\nodesrv,\cprv) \,\,  P_{\cpRV|\nodesRV}(\cprv|\nodesrv) \,\, P_{\nodesRV|\cpRV}(\nodesrv|j).
     \label{eq:transProbZSucc}
\end{equation}

Following similar steps, we can express the transition probabilities from a generic state $(j,\succrv)$ to a state
$(\cprv,0)$ in which the user does not deliver an update as
\begin{align}
     p_{\ancMC}((j,\succrv),(\cprv,0)) =\sum_{\nodesrv=0}^\nodes (1-\pUpdate(\nodesrv,\cprv) )\,\,  P_{\cpRV|\nodesRV}(\cprv|\nodesrv) \,\, P_{\nodesRV|\cpRV}(\nodesrv|j).
     \label{eq:transProbZFail}
\end{align}
It is immediate to verify that the finite-state Markov chain $\ancMCl$ is irreducible and aperiodic, and admits thus a
stationary distribution, which we denote as $\statDistAncMC(\cprv,\succrv)$.

\subsection{Average Peak AoI}
Let us now focus on the calculation of the average peak \ac{AoI} achieved by the frameless ALOHA policy. To this aim, denote by
\peakAoI\ the r.v. describing the value of the \ac{AoI} in stationary conditions at the end of a \ac{CP} over which the
user of interest delivers an update. As exemplified in Fig.~\ref{fig:aoiTimeline}, this quantity can be conveniently
expressed as $\peakAoI=\initAoI+\interUpdate$. Here, $\initAoI$ captures the value at which the \ac{AoI} $\AoI(t)$ was
lastly reset, and corresponds to the duration of the \ac{CP} in which the previous update was received. The r.v.
$\interUpdate$ accounts for the duration of the inter-update time, i.e., the number of slots that have elapsed between
the last and the current successful reception of a status update from the user. The average peak \ac{AoI} can then be
computed as
\begin{equation}
\avPeakAoI = \expOp[\initAoI] + \expOp[\interUpdate].
\label{eq:peakAoI_general}
\end{equation}

Consider first $\initAoI$.
Its \ac{PMF} $P_{\initAoI}(\initaoi)$ can be readily computed from the stationary distribution of the Markov chain $\ancMCl$. We have indeed
\begin{equation}
  P_{\initAoI}(\initaoi) = \frac{\statDistAncMC(\initaoi,1)}{\sum_{\delta=1}^{\maxFrame} \statDistAncMC(\delta,1)}
  \label{eq:pmfDelta0}
\end{equation}
where the numerator denotes the probability for the system to be in a \ac{CP} of duration $\initaoi$ slots in which the
tracked user is decoded, and the denominator is a normalization factor, capturing that we are interested only in
\acp{CP} with successful updates from that user. The first addend in \eqref{eq:peakAoI_general} can be evaluated simply
by applying the definition of expected value.

We analyze now $\expOp[\interUpdate]$.
We start by noting that the statistics of \interUpdate\ depends on the r.v. \initAoI. In fact, the duration of the
\ac{CP} over which the last update was received does influence the number of users contending on the subsequent one,
impacting both the probability for the user of interest to transmit and be decoded as well as the duration of the subsequent \acp{CP}.
It is then convenient to express $\expOp[\interUpdate]$ as
\begin{align}
  \expOp[\interUpdate] = \sum_{\initaoi=1}^{\maxFrame} \expOp[ Y \,|\, \initAoI = \initaoi ] \cdot  P_{\initAoI}(\initaoi).
  \label{eq:interUpdate_general}
\end{align}
Without loss of generality, let us denote by $\ell=1$ the index of the first \ac{CP} that contributed to the inter-update time being tracked. Accordingly, we reformulate the conditional expectation in \eqref{eq:interUpdate_general} considering the value of $\ancMC^{(1)}$ as
\begin{align}
  &\expOp[ \interUpdate \,|\, \initAoI = \initaoi ]  \\[.3em]
  &= \sum_{z} \,\expOp[ \interUpdate \,|\, \ancMC^{(1)} \!\!= z, \initAoI=\initaoi ] \,  \pr\{ \ancMC^{(1)} \!\!= z \,|\, \initAoI=\initaoi \}\\
  &\stackrel{(a)}{=}\sum_{z} \,\expOp[ \interUpdate \,|\, \ancMC^{(1)} \!\!= z] \,  \pr\{ \ancMC^{(1)} \!\!= z \,|\, \initAoI=\initaoi \}
  \label{eq:interUpdate_givenDelta0}
\end{align}
where the summation is taken over all the possible states \mbox{$z = (\cprv,\succrv)$}, $\cprv\in\{1,\dots,\maxFrame\}$,
$\succrv\in\{0,1\}$, and (a) follows from the Markov property of the involved processes. Note that the factors $\pr\{
\ancMC^{(1)} \!\!= z \,|\, \initAoI=\initaoi \}$ on the right-hand side of~\eqref{eq:interUpdate_givenDelta0}
can be computed using~\eqref{eq:transProbZSucc} and~\eqref{eq:transProbZFail}.

The conditional expectation of \interUpdate\ given the outcome of the first \ac{CP} of the inter-update period, $\expOp[
\interUpdate \,|\, \ancMC^{(1)} \!\!= z, \initAoI=\initaoi ]$, can be derived resorting to a first step analysis
\cite{TaylorKarlin}. To this aim, let us denote by $\mathcal A$ the set of states for the Markov chain $\ancMCl$
corresponding to an update delivery for the user of interest:
\begin{align}
  \mathcal A := \{ z= (\cprv,\succrv) \,|\, \cprv \in \{1,\dots,\maxFrame\}, \succrv = 1 \}.
\end{align}
We refer to $\mathcal A$ as the set of \emph{absorbing} states, and define the chain absorption time as
\begin{align}
  \ell^* := \min\{ \ell \geq 1 \, | \, \ancMCl \in \mathcal A\}.
\end{align}
Furthermore, let us assign to each state $z =(\cprv,\succrv)$ a cost
\begin{align}
  g((\cprv,\succrv)) = \cprv.
\end{align}
Then, the overall duration (in slots) of the inter-update period \interUpdate\ is simply given by the sum of the costs
undergone for all states traversed by the process \ancMCl\ from time $\ell=1$ until absorption, i.e.,
\begin{align}
  \expOp[\interUpdate \,|\, \ancMC^{(1)}= z] = \expOp\left[\left. \,\,\sum_{\ell=1}^{\ell^*} g\big(\ancMCl\big)
    \,\right|\,\ancMC^{(1)} =z \,\,\right].
\end{align}
To compute this quantity, consider first the situation in which the packet from the user of interested is decoded
already in the initial \ac{CP}.
In this case, the chain is immediately absorbed, and \interUpdate\ coincides with the length of the initial \ac{CP}:
\begin{align}
\expOp[\interUpdate \,|\, \ancMC^{(1)}= (\cprv,1)] = \cprv.
\label{eq:firstStep_a}
\end{align}
When $\ancMC^{(1)} = (\cprv,0) \not\in \mathcal A$, instead, the average cost prior to absorption can be computed by conditioning on the outcome of the first transition. Specifically,
\begin{align}
  \expOp[\interUpdate \,|\, \ancMC^{(1)}= (\cprv,0)] = \cprv + \sum_{z} \expOp[\interUpdate \,|\, \ancMC^{(1)}= z] \cdot p_{\ancMC}((\cprv,0),z)\\[-2em]
  \label{eq:firstStep_b}
\end{align}
where the Markov property ensures that the average cost once the transition to state $z$ is taken is equal to the one that we would have by starting from such state.
Combining \eqref{eq:firstStep_a} and \eqref{eq:firstStep_b}, we get a full-rank system of $\maxFrame$ equations in the
\maxFrame\ unkowns $\expOp[\interUpdate\,|\,\ancMC^{(1)}=(\cprv,0)]$. We obtain $\expOp[\interUpdate]$ by solving such
system, and by insert the results into
\eqref{eq:interUpdate_givenDelta0} and \eqref{eq:interUpdate_general}.
Finally, the sought average peak \ac{AoI} \avPeakAoI\ follows as per \eqref{eq:peakAoI_general}.

\begin{figure}
    \centering
    \includegraphics[width=.85\columnwidth]{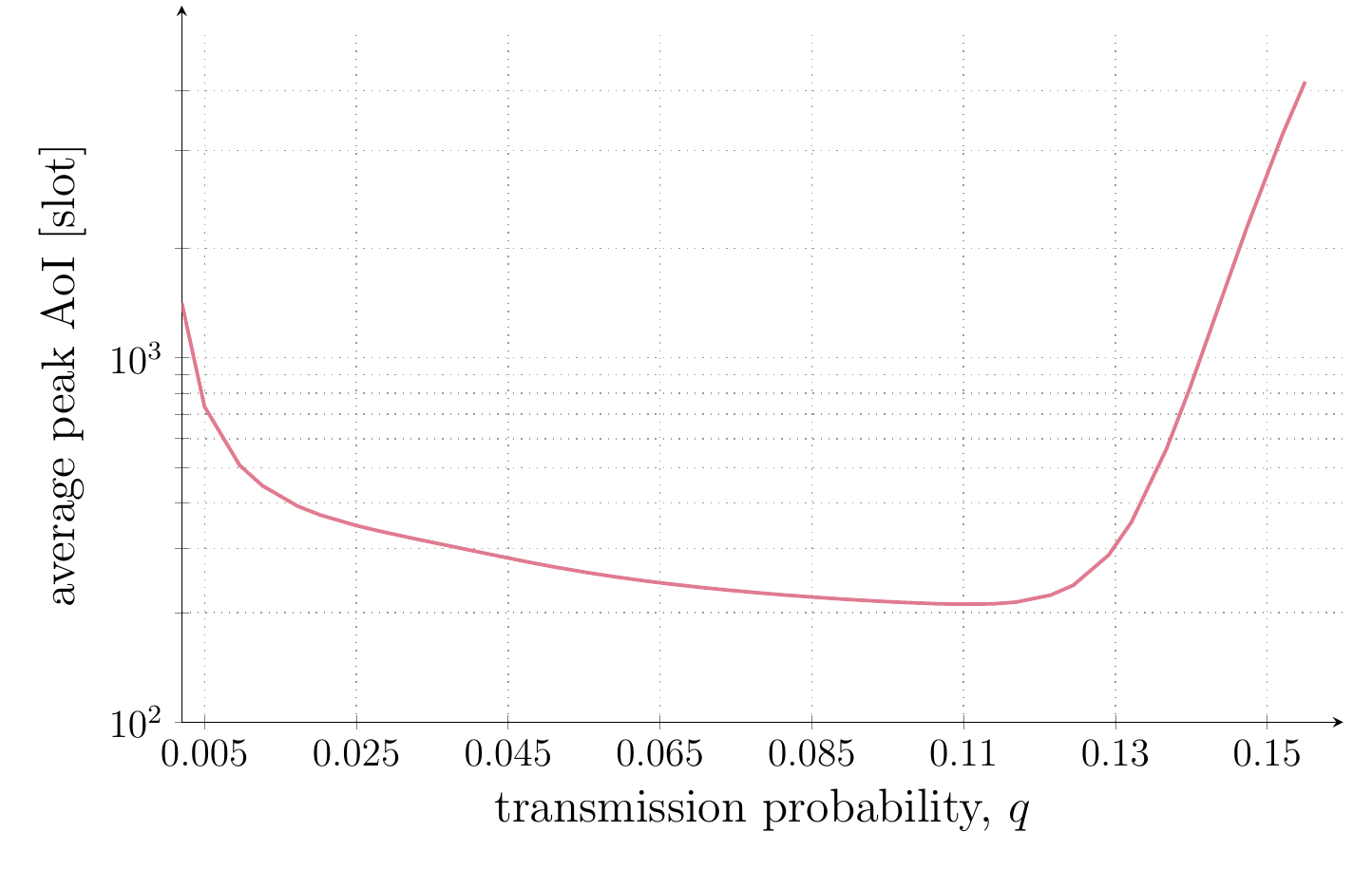}
    \caption{Average peak \ac{AoI} \avPeakAoI\ vs. transmission probability \pTx. Results obtained with $\nodes=100$ users, $\maxFrame=100$ slots. The packet generation probability has been set such that $\pAct\nodes=0.6$.}
    \label{fig:aoiVsPtx}
\end{figure}

In Fig.~\ref{fig:aoiVsPtx}, we plot the average peak \ac{AoI} against the channel access probability \pTx\ for the same
setting considered in Fig.~\ref{fig:truVsPtx}, i.e., $\nodes=100$, $\maxFrame=100$, $\pAct \nodes = 0.6$.
As illustrated in the plot,  both too low and too high values of \pTx\ result in poor performance in terms of \ac{AoI}. In the former case, an excessively conservative behavior is likely to result in a user missing opportunities to
deliver a status update, refraining from transmission for the whole duration of a \ac{CP} even when a packet is
available.
Conversely, collisions dominate when users become too aggressive, hindering the capability of the receiver to decode
transmitted updates prior to reaching the maximum contention duration.

It is also important to note that, for a given traffic profile ($\nodes$, $\pAct$) and a given maximum frame duration, the
optimal operating points in terms of throughput and average peak \ac{AoI} coincide. In other words, there exists a value
$\pTx^{*}$ of the transmission probability that jointly maximizes \tru\ and minimizes \avPeakAoI.
This outcome is common to other random access solutions under symmetric traffic conditions, as epitomized by the
inverse proportionality of \ac{AoI} and throughput exhibited by slotted ALOHA \cite{Yates17:AoI_SA,Munari21_TCOM}. From
this standpoint, indeed, any choice of $\pTx\neq \pTx^*$ reducing the probability to deliver a status update would also
be harmful in terms of information freshness.

\begin{figure}
    \centering
    \includegraphics[width=.85\columnwidth]{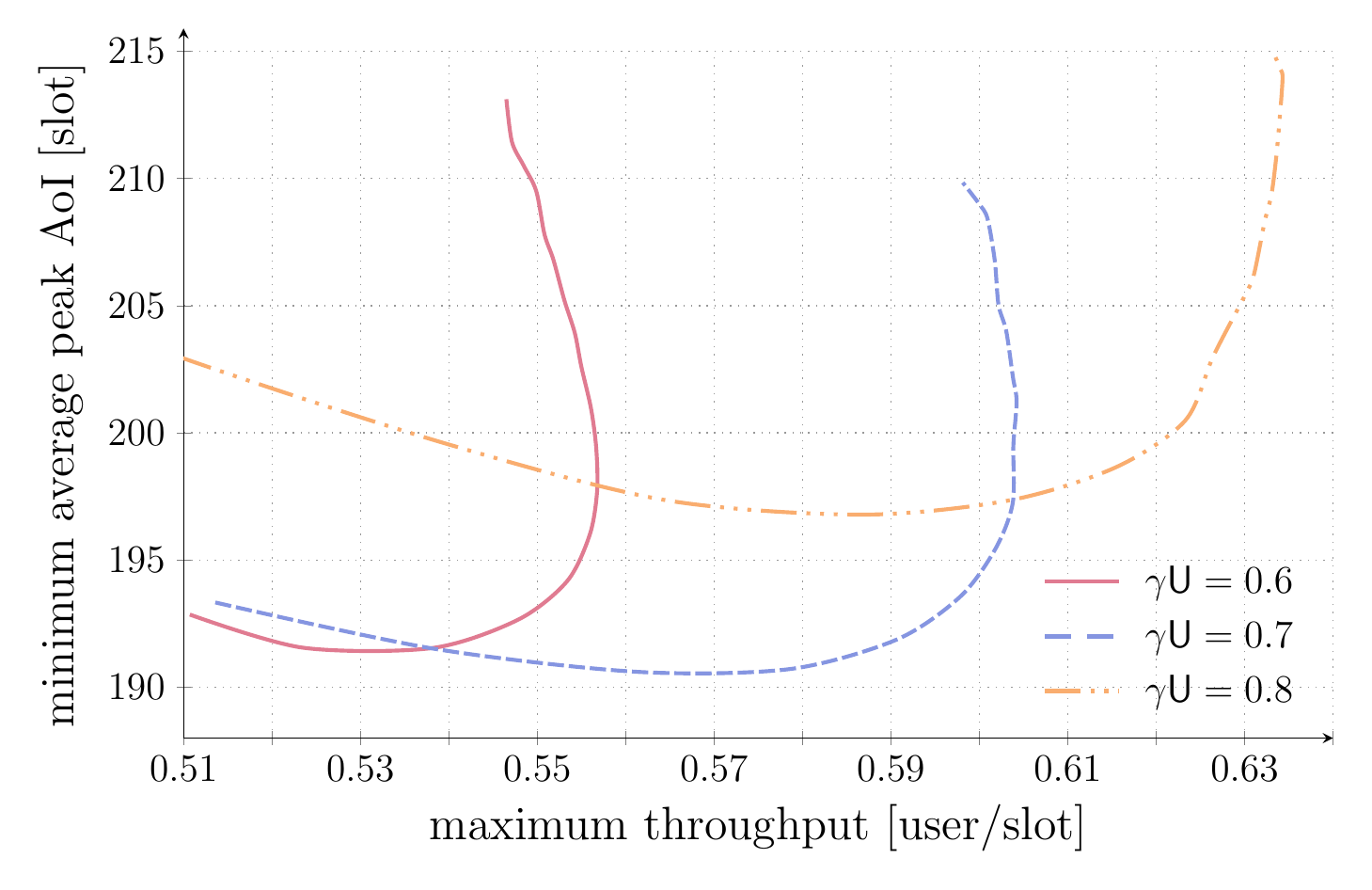}
    \caption{Maximum throughput and minimum average peak \ac{AoI} obtained when varying the maximum \ac{CP} duration \maxFrame\ in the range $\{10,150\}$ slots. In all cases, $\nodes=100$ users were considered.}
    \label{fig:contourTruAge}
\end{figure}

In contrast, frameless ALOHA exhibits a more complex behavior when performance are analyzed versus \maxFrame.
In Fig.~\ref{fig:contourTruAge}, we report the optimal
throughput and peak \ac{AoI} pairs that can be achieved by tuning the maximum \ac{CP} duration. Specifically, we explore
different values of \maxFrame\ between $10$ and $150$ slots, and pick, for each setting, the optimal access probability
$\pTx^*$, plotting the corresponding values obtained for \tru\ and \avPeakAoI. Distinct curves in the figure refer to
different packet generation probabilities. In all cases, $\nodes=100$.

Consider first the case $\pAct\nodes=0.6$, represented by the solid curve in Fig.~\ref{fig:contourTruAge}, and focus on
throughput performance. For low values of \maxFrame, the system operates in the lower-left corner of the plot. Too short
\acp{CP} hinder packet decoding, not allowing enough slots for \ac{SIC} to be fully efficient. By increasing
the maximum duration of the \ac{CP}, \tru\ improves, approaching the elbow exhibited by the reported trend. After a
certain point, though, a further increase of \maxFrame\ enables to decode only a limited additional number of users, and
such diminishing-return behavior leads to a decrease in throughput.

Notably, while a similar trend emerges also for the average peak \ac{AoI}, the impact of operating over excessively long
\acp{CP} is far more pronounced. The rationale behind this lies in the dependency of \avPeakAoI\ on the inter-update
time, i.e., the number of \acp{CP} between two updates as well as their duration in slots. From this standpoint, higher
values of \maxFrame\ may reduce the former (increasing throughput), yet entail a larger average cost in terms of elapsed
slots over a \ac{CP}. While initially the first factor prevails, and \avPeakAoI\ improves together with \tru, the
impact of longer \acp{CP} quickly emerges and yields a reduction of \avPeakAoI.


This trend is confirmed when analyzing the results obtained by increasing the channel load $\pAct\nodes$.  As shown in
Fig.~\ref{fig:contourTruAge}, frameless ALOHA can reach larger values of throughput by supporting higher levels of
channel congestion.
Yet operating in this region is detrimental in terms of average peak \ac{AoI}.


%% file: conclusions.tex
\section{Conclusions}
\label{sec:conclusions}
We provided a characterization of the average throughput and the average peak \ac{AoI} achievable in a random access
system operating according to the frameless ALOHA protocol.
One novel aspect of our analysis, compared to results available in the literature, is that we are able to capture the
rich system dynamics that result from letting the number of users contending on a given \ac{CP} depend on the duration
of the previous \ac{CP}.
We show that this modeling assumption is critical to capture practically relevant trade-offs in the design of the
system.
Specifically,  choosing the maximum length of the \ac{CP} so as to
maximize the system throughput may result in a significant deterioration of the average peak \ac{AoI}.
We hasten to add that the performance of frameless ALOHA protocol considered in the paper can be further improved by
making some of the protocol parameters dependent on the duration of the previous \ac{CP}.
Such a generalization will be addressed in a future work.